\documentclass[a4paper,11pt]{article}
\usepackage[utf8]{inputenc}
\usepackage{color}
\usepackage{xcolor}
\usepackage{amsmath}
\usepackage{amsthm}
\usepackage{adjustbox}
\usepackage{graphicx}
\usepackage{float}
\usepackage{multirow}
\usepackage[mathscr]{eucal}
\usepackage{subfigure}
\usepackage{epstopdf}
\usepackage{caption}
\usepackage{verbatim}
\usepackage{natbib}
\usepackage{colortbl}
 	\definecolor{manatee}{rgb}{0.59, 0.6, 0.67}

\usepackage{ulem}

\makeatletter
\newcommand*\bigcdot{\mathpalette\bigcdot@{.7}}
\newcommand*\bigcdot@[2]{\mathbin{\vcenter{\hbox{\scalebox{#2}{$\m@th#1\bullet$}}}}}
\makeatother

\title{Bayesian estimation for conditional probabilities associated to directed acyclic graphs: study of hospitalization of severe   influenza cases}

\author{Lesly Acosta$^{a}$, Carmen Armero$^{b}$   \vspace*{0.2cm}\\
\small $^{a}$Department of Statistics and Operations Research, Universitat Polit\`ecnica\\ \small de Catalunya, Barcelona-TECH\\
        \small $^{b}$Department of Statistics and Operations Research, Universitat\\
        \small de València, Burjassot, Spain \\\\
        \small $^{*}$Corresponding author: Lesly Acosta; \tt{lesly.acosta@upc.edu}
}

\begin{document}
\maketitle

\thispagestyle{empty}


\begin{abstract}
This paper presents a Bayesian inferential framework for estimating joint, conditional, and marginal probabilities in directed acyclic graphs (DAGs)  applied to the study of the progression of hospitalized patients with severe influenza. Using data from the PIDIRAC retro\-spective cohort study in Catalonia, we model patient pathways from admission through different stages of care until discharge, death, or transfer to a long-term care facility. Direct transition probabilities are estimated through a Bayesian approach combining conjugate  Dirichlet-multinomial inferential processes, while posterior distributions associa\-ted to absorbing state or inverse probabilities are assessed via simulation techniques. Bayesian methodology quantifies uncertainty through posterior distributions, providing insights into disease progression and improving hospital resource planning during seasonal influenza peaks. These results support more effective patient management and decision-making in healthcare systems.\\

\noindent \textbf{Keywords:} Confirmed influenza hospitalization;   Directed acyclic graphs (DAGs);   Dirichlet-multinomial Bayesian inferential process;   Healthcare decision-making; Transition probabilities.
\end{abstract}


\renewcommand{\baselinestretch}{1.}
\bigskip

\section{Introduction}

According to the World Health Organization (WHO): ``Seasonal influenza (the flu) is an acute respiratory infection caused by influenza viruses common in all parts of the world''.  Following official estimates, about 1 billion cases of seasonal influenza occur worldwide each year. This includes between 3 and 5 million cases of severe illness and between 290 and 650 thousand respiratory deaths caused by the disease   \citep{p23}.
Influenza rates are very high in children but its mortality records are shocking in elderly populations as well as in people affected by chronic diseases. It is a worldwide cause of hospital admissions and mortality in these latter groups   \citep{macias}.

Influenza prevention and control remains a serious public health challenge, despite the availability of vaccines and antiviral treatments \citep{cs}. The European Center for Disease Prevention and Control (ECDC) is an agency of the European Union (EU) that collects epidemiological and virological data from member countries of the European Economic Area (EEA). Surveillance data   come from the sentinel influenza surveillance systems of each associated country, which may cover substantial parts of the population or even have a universal surveillance system \citep{sb}.

\enlargethispage{1cm}
Although most people affected by the seasonal influenza recover within one or two weeks without medical attention, it can cause serious illnesses and mortality, especially among population at higher risk.  Severe influenza complications can result in hospitalization, possibly with admission to the ICU or even death (\cite{lavir21}, \cite{sa21}). According to ECDC, around 10 to 30$\%$  of Europe's population is infected annually with influenza, causing hundreds of thousand hospitalizations. A systematic review of the clinical burden of influenza disease in older people was done by \cite{LJal23} with data from January 2012 to February 2022.

There are almost no studies that quantitatively analyse the different health conditions that   hospitalised patients with severe influenza  can experience from admission to discharge.   Knowledge of these pathways would be a valuable tool for improving hospital resource planning and organization of the seasonal period of influenza, the winter. In Europe, influenza generally causes annual epidemics that affect up to 20$\%$   of the population.

Graph theory is a very theoretical mathematical subject  with an enormous power to visualize the basic functioning of scenarios that operate in environments with many sources  of uncertainty. In particular, the   evolution of a hospitalized patient from admission to discharge can be represented graphically by means of a probabilistic directed acyclic graph (DAG) (\cite{Cowell}, \cite{Barber}) with nodes defined by the different health conditions of the inpatients and arrows connecting   two consecutive nodes without any possibility of return. Transition probabilities between  nodes are conditional probabilities  that provide valuable clinical information on the state of health in which individuals move from their current status. They are the basis for assessing the uncertainty associated with the different trajectories of the study, the final (absorbing) states, and inverse probabilities that inform previous events.

This paper presents a general inferential procedure for estimating joint, conditional, and marginal probabilities associated to random events in probabilistic DAGs, which we apply to   assess the different pathways that a patient with severe influenza may follow from their admission to the hospital to their discharge, as fully cured, dead or sent to a long-stay facility.  These la\-tter type of institutions usually welcome patients with chronic diseases and comorbidities who have little hope of cure, but require continued clinical care.

All the inferential processes in this work are framed within the Bayesian inferential methodology. This will allow to directly quantify the uncertainty associated with the relevant outcomes through probability distributions. In particular, posterior distribution associated to transition probabilities or absorbing states  will allow us to better understand the hospital evolution of patients with severe influenza. Overall, they may be a useful tool in the effective management of patients hospitalised with influenza during peaks of influenza epidemic activity.

This paper is organized as follows. Section 2 presents a description of the motivated problem, named the PIDIRAC cohort study,  and data as well as the DAG that represents the evolution  of a severe influenza patient in hospital. Section 3 introduces the general
Bayesian  modeling for assessing   conditional probabilities for adjacent and non adjacent states of a DAG.
   Sec\-tion 4 applies
the general approach in Section 3 to the data of the PIDIRAC study.   Finally, Section 5 contains  some conclusions.

\section{PIDIRAC retrospective cohort study}

This study focuses on hospitalized  severe influenza patients.  Data for the analysis were collected from a retrospective cohort study of hospitalized, laboratory confirmed, influenza (SHLCI) patients registered from 1 October 2017 to 22 May 2018 by the 14 hospitals included in the Primary Care Influenza Surveillance System of Catalonia (PIDIRAC). Catalonia  is an autonomous community of Spain   located in the northeast of the Iberian Peninsula with a population of around 8 million inhabitants. The median and IQR age of the  hospitalized patients was 72 and 59$-$83 years respectively, with 563 (43$\%$) of all  being female.

All severe influenza patients who came to the hospital were initially attended by a physician who, depending on the patient's state of health, recommended admission to an intensive care unit (ICU, $I$ from now on) or to a specific hospital ward ($W1$). Some of the patients who were initially directed to $W1$ were later transferred to the ICU, derived to a long-term care facility ($L$), died ($D$) or were cured and discharged from the hospital and sent home ($H$).  Patients in ICU can die or, if they improved, be sent to a second type of ward ($W2$) of the hospital, from where they can move to $H$, $L$ or $D$.

\begin{figure}[H]
\begin{center}
\includegraphics[width=14.1cm]{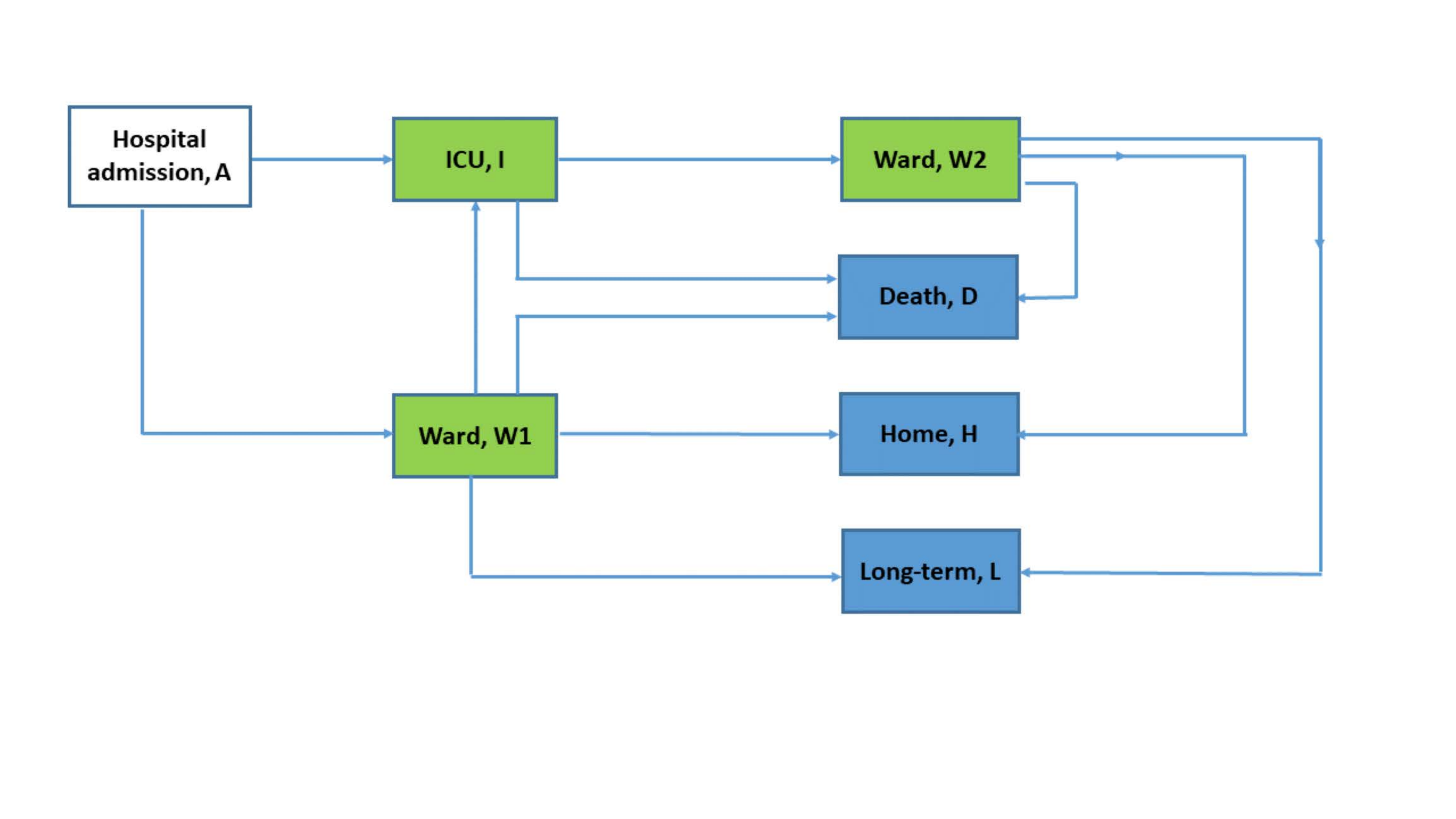} \vspace*{-1.5cm}
\caption{Directed acyclic graph for a description of the progress of patients with severe influenza in hospital. Transient states in green and absorbing states in blue.}\label{figure1}
\end{center}
\end{figure}

Consider a graph $G=(S,\,\mathcal C)$ with nodes $S=\{A,\, I, \,W1, \,W2, \,D, \,H, \,L\}$   the different states or services in the hospital and ordered arcs $$\mathcal C=\{A\,I, \,A\,W1, \,W1\,I, I\,W2,\,W1\,D, \,W1\,H, \,W1\,L, \,W2\,D, \,W2\,H, \,W2\,L\}$$ connecting the different adjacent nodes. See Figure \ref{figure1} for a DAG that re\-presents the evolution of hospitalized patients with severe influenza through nodes describing their different health states and/or hospital services used and the different directed arcs connecting neighboring nodes. The nature of the different states is different: The admission of a patient to the hospital is the initial state $A$, states $I$, $W1$, and $W2$ are transient, and states $D$, $H$, and $L$ absorbing.

  Over the study period, the number of patients with a diagnosis of severe influenza  admitted to the hospital was  1$\hspace{0.01cm}$306. Of these patients,  1$\hspace{0.01cm}$208   were initially referred to the ward and 98 were sent to the ICU. A total of 82 patients in $W1$ were subsequently transferred to the ICU. Of the total of  1$\hspace{0.01cm}$226 patients on the ward not sent to the ICU, 946 were discharged and sent home, 55 were sent to $L$, and 125 died. A total of 35 patients died during their stay in ICU; and the rest, 145, were sent to $W2$ from where 118 were discharged and sent home, 12 were sent to a long-stay facility and 15 died.

\section{Bayesian modeling of conditional probabilities }

The  probabilistic approach to our model is concentrated on conditional pro\-babilities that assess the uncertainty associated to  the different paths between the states of the process: probabilities of visiting a certain state from ano\-ther given state, directly without intermediate states or not; probabilities of ending in each of the absorbing states; and even, inverse probabilities that focus on assessing the uncertainty associated with a previous state knowing that the process has ended up visiting a certain state subsequently.

\subsection{Dirichlet-multinomial learning process for assessing direct conditional  probabilities}

Assume a graph with a finite set $S$ of states and consider (without loss of generality) that from the state $i \in S$    it is possible to  directly visit the states  $\{1,\ldots,J\} \in S$ with probabilities $\boldsymbol \theta_{i}=(\theta_{i1}, \ldots, \theta_{iJ})'$, where $\theta_{ij}$, $i \neq j$,  represents the probability that the individual's departure from state $i$  will be to visit state $j$ directly, with $\theta_{iJ}=1-\mbox{$\sum$}_{j=1}^{J-1} \theta_{ij}$.

Let $\boldsymbol Y_{i}=(Y_{i1}, \ldots, Y_{iJ})'$ be the random multinomial vector whose component $j$, $Y_{ij}$, describes the number of individuals that move directly from state $i$  to state $j$ from the $n_i$ individuals that are in the state $i$ and eventually leave it, where $Y_{iJ}=n_i-\mbox{$\sum$}_{j=1}^{J-1} \, Y_{ij}$. The parametric vector associated to this multinomial distribution is $\boldsymbol \theta_{i}$.

The basic Bayesian inferential process for multinomial  probabilities $\boldsymbol \theta_i$ account for  the Dirichlet family of distributions as the conjugate family for the sampling multinomial distribution, $\boldsymbol Y_i \mid \boldsymbol \theta_{i} \sim \mbox{Mn}(n_i, \boldsymbol \theta_{i})$. This implies that if a Dirichlet distribution $\mbox{Di}(\boldsymbol \alpha_i)$ with parameters $\boldsymbol \alpha_i=(\alpha_{i1},  \ldots,  \alpha_{iJ})$  is chosen as a prior distribution
$\pi(\boldsymbol \theta_i)$  for $\boldsymbol \theta_i$,  the corresponding posterior distribution $\pi(\boldsymbol \theta_i \mid \mathcal D)$, where $\mathcal D$ stands for the data,  will also be a Dirichlet distribution,  $\mbox{Di}( \boldsymbol y_i + \boldsymbol  \alpha_i )$, where $\boldsymbol y_i$ is the vector of the observations from $\boldsymbol Y_i$. Just as the binomial distribution is the univariate version of the multinomial distribution, the Beta distribution is the univariate version of the Dirichlet distribution. In this regard, the posterior marginal distribution associated to probability $\theta_{ij}$, $\pi(\theta_{ij} \mid \mathcal D)$,  is a Beta distribution with parameters
$y_{ij}+\alpha_{ij}$ and $\alpha_i^{+}+n_{i}-(y_{ij}+\alpha_{ij})$, where $\alpha_i^{+}=\mbox{$\sum$}_{j=1}^J \, \alpha_{ij}$   \citep{congdon05}.

There are many studies where prior information cannot be provided or is available but it is not intended to be used in the inferential process in order to make the analysis more ``objective''. In these scenarios, the Bayesian inferential protocol establishes as necessary the specification of   a prior distribution that serves as a starting point for the inference and that disturbs and distorts as little as possible the information provided by the data. The choice of such  scarce informative distribution  is not unanimous in the scientific literature and has generated a lot of controversy. We will not go into this issue and will use the Perk's prior Dirichlet   distribution (Perks, 1947; Berger et al., 2015), whose parameters share the unit of probability equally and is  the most widely used Dirichlet distributions for this type of problems. The corresponding   posterior distribution   is $\mbox{Di}(\boldsymbol y_i + \boldsymbol 1/J )$, where $\boldsymbol 1$ is now a vector of ones of dimension $J$.

\subsection{Bayes inference for non-direct conditional probabilities}

A little more notation should be added to the study to analyse conditional probabilities associated with non-contiguous states. We define $\theta_{ik_{1}\ldots k_{K}j}$ as the probability associated with the path that begins in state $i$, then visits state $k_1$, immediately after  state $k_2$, and so on up to $k_K$, and finally ends in $j$, where each of the visited states is temporally following its immediate preceding. In the event the state $i$ is temporally earlier than state $j$,   $\theta_{i   \bigcdot   j}$  shall stand for the probability of visiting $j$ from $i$ via any path $\mathcal P(i,j)$ that connects both states, $\theta_{i \bigcdot j} = \mbox{$\sum$}_{\mathcal P(i,j)}\, \theta_{\mathcal P(i,j)}$. Analogously, if state  $i$ is temporally later than state $j$,   $\theta_{i \bigcdot j}$ will represent the probability of having departed from state $j$ knowing that the process has visited  the posterior state $i$.  This probability  is calculated from Bayes's theorem as follows
$$ \theta_{i \bigcdot j} = \frac{ \theta_{j \bigcdot i} \, \theta_{ i}}  {\theta_{ j}} ,  $$
 \noindent where now $\theta_i$ indicates the probability of visiting state $i$ from the initial state of the process, $\theta_{A \bigcdot I}$.

 The  conjugate Dirichlet-multinomial learning process is suitable for the computation of   posterior  distributions associated to   jumping probabilities between   a state and its immediate next states, but not for posterior distributions associated to probabilities between non-neighboring states or inverse probabilities. In this sense, we will assume   a Markovian structure for the transition probabilities as follows:
 $$\theta_{ik_{1} \ldots k_{K}j} =\theta_{ik_{1}}\,\theta_{k_{1}k_{2}}.\ldots\theta_{k_{K}j}.$$
This condition allows the simulation of  the posterior distribution  $\pi(\theta_{i \bigcdot j} \mid \mathcal D)$ of the probability associated to any path connecting  states $i$ and $j$ of the process. In particular, if  $i$ is a state earlier than $j$ and we consider the trajectory from $i$ to $j$ through the consecutive states $k_1, k_2, \ldots, k_K$, a random sample  from the posterior distribution $\pi(\theta_{ik_{1} \ldots k_{K}j} \mid \mathcal D)$ is
$\{\theta_{ik_{1} \ldots k_{K}j}^{(m)}=\theta_{ik_{1}}^{(m)}\,\theta_{k_{1}k_{2}}^{(m)}.\ldots\theta_{k_{K}j}^{(m)}, \,m=1, \ldots, M \}$, where each simulated value
$\theta_{k_ik_j}^{(m)}$  is generated from the subsequent  Beta marginal posterior distribution, $\pi(\theta_{ k_{i} \, k_{j}} \mid \mathcal D)$. In this way, we can compute a random sample from $\pi(\theta_{i \bigcdot j} \mid \mathcal D)$ taking into account that
\begin{equation}
    \pi(\theta_{i \bigcdot j} \mid \mathcal D) =   \pi(  \mbox{$\sum$}_{\mathcal P(i,j)}\, \theta_{\mathcal P(i,j)} \mid \mathcal D). \label{eqn:1}
\end{equation}
In the case that $i$ is subsequent to $j$, we can generate a random sample from
$\pi(\theta_{i \bigcdot j} \mid \mathcal D)$ bearing in mind that
\begin{equation}
\pi(\theta_{i \bigcdot j} \mid \mathcal D) =   \pi \Big(  \frac{ \theta_{j \bigcdot i} \, \theta_{ i}}  {\theta_{ j}} \mid \mathcal D  \Big ).
\label{eqn:2}
\end{equation}

\section{Hospitalization of severe influenza cases}

\subsection{Probabilities associated with direct transitions between states}

We begin the inferential process of the PIDIRAC  study by estimating the probability distribution associated with visiting each of the states that can be accessed from an immediately preceding state. In our research these would be the random vector  $\boldsymbol \theta_A=(\theta_{A\,W1}, \theta_{A\,I})'$ that assesses the probabi\-lity that a patient admitted to the hospital will be referred to  ward $W1$ or ICU,  $\boldsymbol \theta_I=(\theta_{I\,W2}, \theta_{I\,D})'$ that accounts for the probability that a patient in the ICU moves to ward $W2$ or dies, $\boldsymbol \theta_{W1}=(\theta_{W1\,I}, \theta_{W1\,D},  \theta_{W1\,H},  \theta_{W1\,L} )'$ as the vector that indicates the possible visits to $I$, $D$, $H$ and $L$ from $W1$, and the vector $\boldsymbol \theta_{W2}=(\theta_{W2\,D},   \theta_{W2\,H},\theta_{W2\,L})'$    formed by the probability that a patient ends up in $D$, $H$ or $L$   since his second stay on the ward, $W2$. In all these cases, we will use the Perk prior distribution introduced above as the  non-informative prior distribution of all the inferential processes  in our study.\\

\noindent \textbf{From admission to the ICU or ward} \vspace*{-0.4cm}\\
\enlargethispage{1cm}

 \noindent Data for learning about   the probability $\boldsymbol \theta_A=(\theta_{A\,W1}, \theta_{A\,I})'$ that   a patient admitted to the hospital will be referred to   ward $W1$ or ICU, respectively, refer to the   number of patients admitted to the hospital, 1306, and how many of them were sent to the ward $W1$ or ICU, 1208 and 98, respectively.  The posterior distribution of  $\boldsymbol \theta_A$  is the Dirichlet distribution
$$\pi( \boldsymbol \theta_A \mid \mathcal D)= \mbox{Di}(1208.5, 98.5), $$
 \noindent with posterior expectations  0.925 and  0.075 for $\theta_{A\,W1}$ and
 $\theta_{A\,I}$, respectively, thus showing that about the 92.5$\%$ of hospital patients are transferred directly to $W1$, and the rest, around 7.5$\%$, are sent directly to the ICU. Posterior 95$\%$ credible intervals for these probabilities are $(0.910, 0.938)$  and $(0.062, 0.090)$, respectively. These are very narrow intervals indica\-ting very little uncertainty about both probabilities.  It is interesting to note that these credible intervals provide a direct measure of the uncertainty of
 $\theta_{A\,W1}$ and $\theta_{A\,I}$ that is not possible to consider in the frequentist framework. Figure 2 shows the marginal beta posterior distribution associated  for each of the two probabilities, $\pi(\theta_{AI} \mid \mathcal{D})=\mbox{Be}(98.5, \,1208.5)$ and $\pi(\theta_{A\,W1} \mid \mathcal{D})=\mbox{Be}(1208.5, \,98.5)$. \\

 \begin{figure}[H]
\begin{center}
\includegraphics[width=12.5cm, height=6cm]{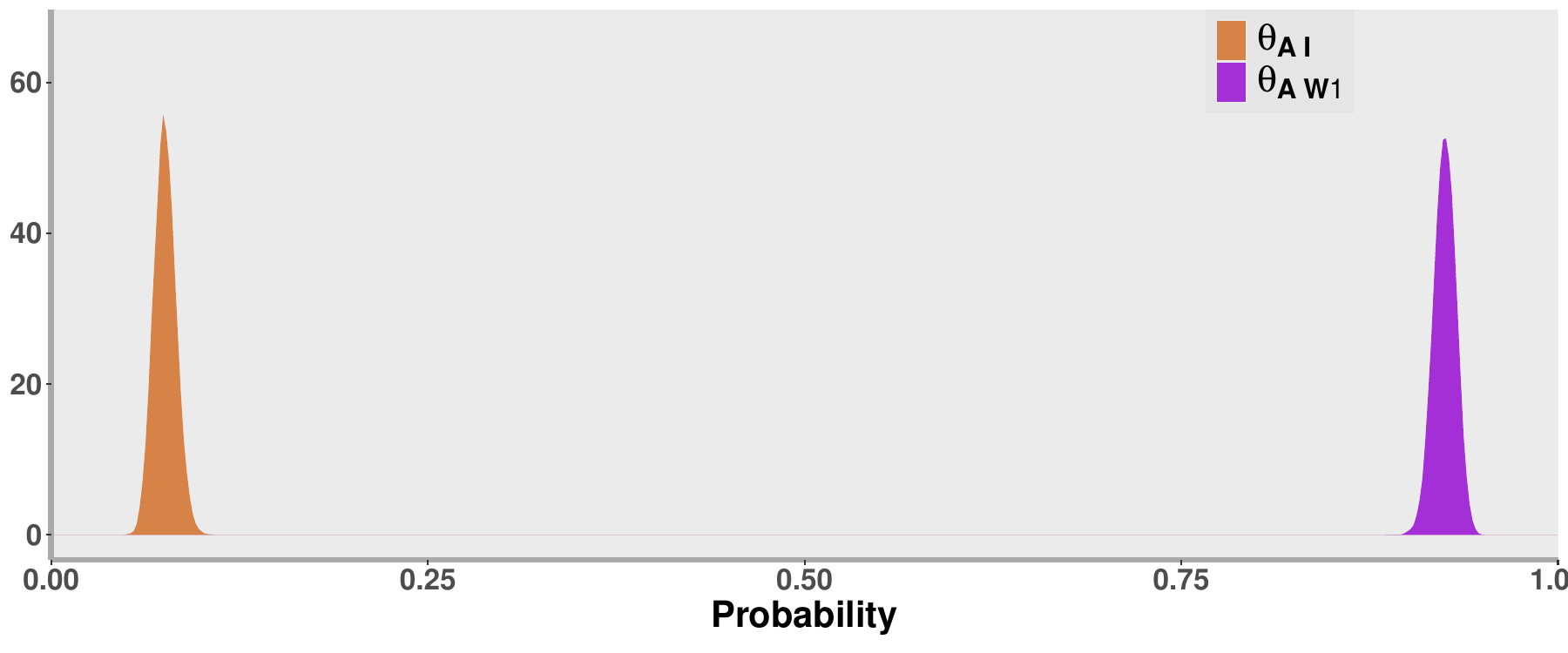}
\caption{Posterior marginal distribution for the probability of a patient admitted to the hospital being sent to ICU, $\theta_{A\,I}$, or ward $W1$,    $\theta_{A\,W1}$.} \label{fig1}
\end{center}
\end{figure}

\noindent \textbf{From   ward $\boldsymbol W1$ to ICU, death, home or a long-stay facility.}\vspace*{-0.4cm}\\

\noindent Of the 1208 patients who were initially sent to   ward $W1$, 82 were transferred to ICU, 946  discharged  home, 55  derived to a long-term care facility $L$, and 125 died. This information generates the following posterior distribution for  $\boldsymbol \theta_{W1}=(\theta_{W1\,I}, \theta_{W1\,D},  \theta_{W1\,H},  \theta_{W1\,L} )'$,  the vector of the probability associated to a possible visit from $W1$ to $I$, $D$, $H$ and $L$, respectively
$$\pi( \boldsymbol \theta_{W1} \mid \mathcal D)= \mbox{Di}(82.25, 125.25, 946.25, 55.25), $$
\noindent with posterior expectation and 95$\%$ credible interval for the posterior marginal distribution associated with each one of the probabilities in
$\boldsymbol \theta_{W1}$ in Table~\ref{tab:1}.

\begin{table}[h!]
\begin{center}
\begin{tabular}{clcc}
Probability & Marginal & Mean & 95$\%$ CI \\
\hline
$\theta_{W1 \,I}$  & Be(82.25, 1126.75)  & 0.068  &   (0.035, 0.058)  \\
$\theta_{W1 \,D}$  & Be(125.25, 1083.75)  & 0.103 & (0.087, 0.121)\\
$\theta_{W1 \,H}$  & Be(946.25, 262.75)  & 0.783 &  (0.759, 0.806)\\
$\theta_{W1 \,L}$  & Be(55.25, 1153.75)  & 0.046 &(0.035, 0.058) \\
\hline
\end{tabular}
\caption{Posterior marginal distribution, mean, and 95$\%$ credible interval for the probabilities in $\boldsymbol \theta_{W1}$.}
    \label{tab:1}
\end{center}
\end{table}

 \begin{figure}[H]
\begin{center}
\includegraphics[width=12.5cm, height=6cm]{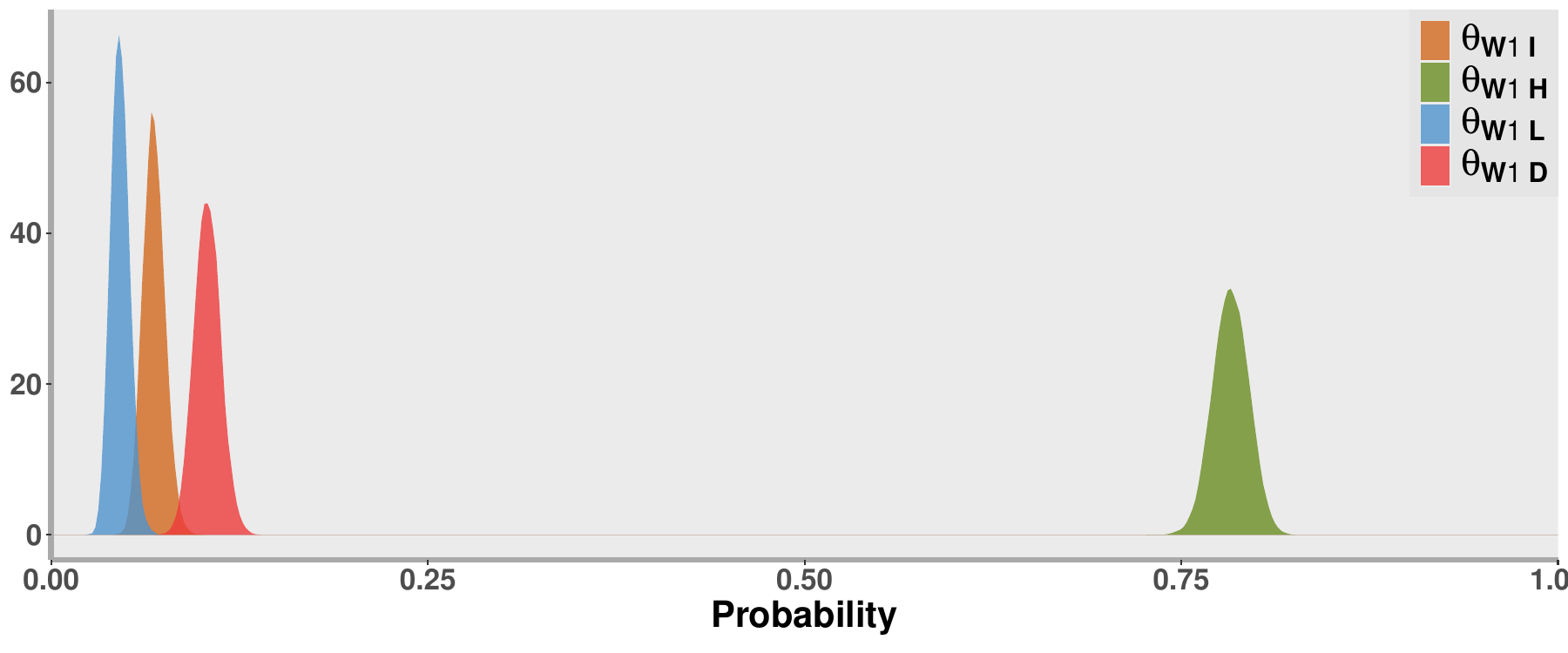}
\caption{Posterior marginal distribution for the probability   of a patient in $W1$ is sent to a long stay facility, $\theta_{W1\,L}$, to ICU,  $\theta_{W1\,I}$, to
death, $\theta_{W1\,D}$, or home, $\theta_{W1\,H}$.}
\label{fig:3}
\end{center}
\end{figure}

Figure \ref{fig:3} shows the marginal posterior distribution of each of the probabilities associated with $\boldsymbol \theta_{W1}$. It is interesting to note the homogeneity of the distributions associated with moving to states $I$, $D$ and $L$ and the difference in magnitude and variability of that associated with $H$. Approximately 10$\%$ of the patients in the ward $W1$ die,  5$\%$ are sent to the ICU, and around 4$\%$ are transferred to a long-term institution, probably with very little chance of cure. On the other hand, about 78$\%$ of patients admitted to the ward are discharged, although the uncertainty associated with this probability is greater than the previous three. \\

\noindent \textbf{From ICU to death or to a second hospital ward, $W2$} \vspace*{-0.4cm}\\

Among the 180 patients who went through the ICU 35 died, and the rest, 145, were referred to $W2$.  As a result, the   posterior distribution for $\boldsymbol \theta_I=(\theta_{I\,W2}, \theta_{I\,D})'$ that accounts for the probability that a patient in the ICU moves to ward $W2$  or dies  is as follows,

$$\pi( \boldsymbol \theta_{I} \mid \mathcal D)= \mbox{Di}(145.50, 35.50), $$

\noindent with posterior expectation 0.804 and 95$\%$ credible interval (0.743, 0.858) for $\theta_{I\,W2}$, and 0.196 and (0.142, 0.257) for $\theta_{ID}$, respectively.  Figure \label{fig:4} shows both posterior distributions, very different in their location but similar in shape and variability.

\begin{figure}[H]
\begin{center}
\includegraphics[width=12.5cm, height=6cm]{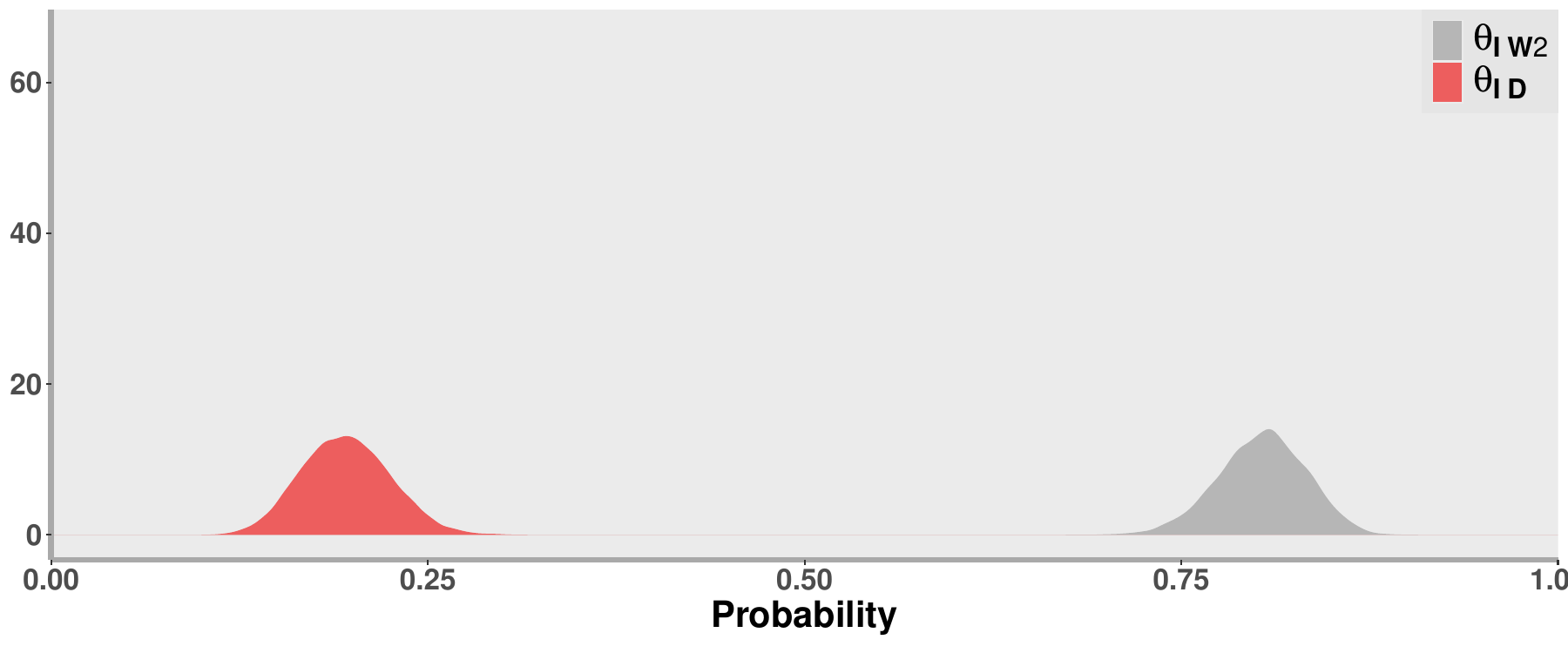}
\caption{Posterior marginal  distribution of the probability that a patient leaving ICU is transferred to  ward, $\theta_{I\,W2}$ or dies, $\theta_{I\,D}.$}\label{fig:4}
\end{center}
\end{figure}

\noindent \textbf{From ward $\boldsymbol \theta_{W2}$ to death, home or a long-term care facility} \vspace*{-0.4cm}\\

 A total of 145 patients passed through ward $W2$. Of these, 15 died, 118 were discharged and sent home, and the remaining 15 were sent to a long-term care facility. With this information, the posterior distribution associated to  $\boldsymbol \theta_{W2}=(\theta_{W2\,D},   \theta_{W2\,H},\theta_{W2\,L})'$  is:

$$\pi( \boldsymbol \theta_{W2} \mid \mathcal D)= \mbox{Di}(145.33, 118.33, 12.33),$$

\noindent   with posterior mean and 95$\%$ credible interval for the posterior marginal distribution associated to each one of the probabilities in $\boldsymbol \theta_{W2}$  given in Table~\ref{tab:2}.

\begin{table}[h!]
\begin{center}
\begin{tabular}{clcc}
Probability & Marginal & Mean & 95$\%$ CI \\
\hline
$\theta_{W2 \,D}$  & Be(145.33, 130.67)  & 0.105  &   (0.061, 0.159)  \\
$\theta_{W2 \,H}$  & Be(118.33, 157.67)  & 0.810 & (0.743, 0.870)\\
$\theta_{W2 \,L}$  & Be(12.33, 263.67)  & 0.085 &  (0.045, 0.135) \\
\hline
\end{tabular}
\caption{Posterior marginal distribution, mean and 95$\%$ credible interval for the probabilities in $\boldsymbol \theta_{W2}$.}
    \label{tab:2}
\end{center}
\end{table}

Approximately, 81$\%$  of the  patients leaving $W2$ are discharged and sent home, 10$\%$ die, and the rest, about 9$\%$, are sent to a long-term care facility.

\begin{figure}[H]
\begin{center}
\includegraphics[width=12.5cm, height=6cm]{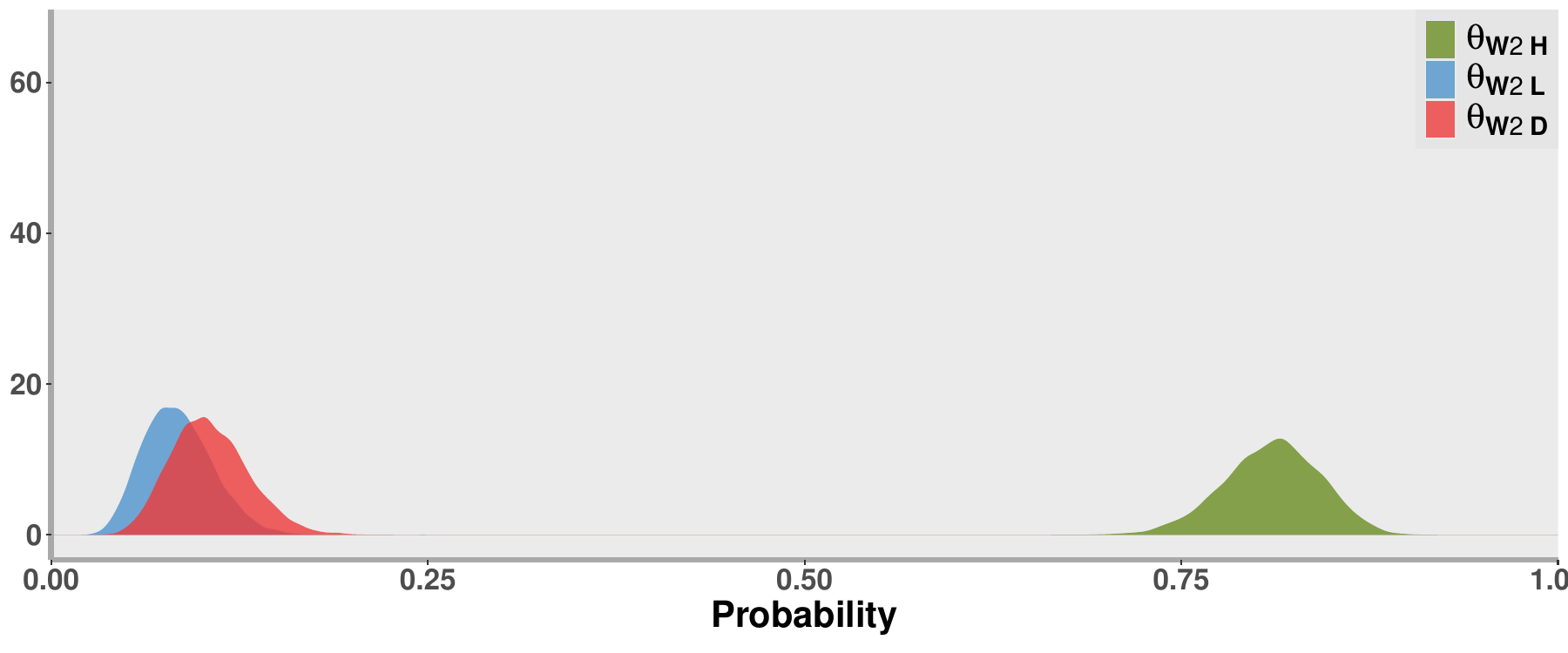}
\caption{Posterior marginal distribution of the probability that a patient in ward $W2$ dies, $\theta_{W2\,D}$, is sent home with a discharge, $\theta_{W2\,H}$, or to a long-stay facility, $\theta_{W2\,L}$. }\label{fig:5}
\end{center}
\end{figure}

Figure \ref{fig:5} shows the corresponding Beta marginal distributions: very different in location, but with very similar variability and shape.

\subsection{Probability of leaving hospital on discharge,
dying, or being sent to a long-term care facility.}

From a clinical point of view, it is important to assess the probability that a patient who enters the hospital with severe influenza will eventually be discharged home, $\theta_{A \bigcdot H}$, die, $\theta_{A \bigcdot D}$, or be sent to a long-term care facility, $\theta_{A \bigcdot L}$. These terminal probabilities are defined in terms of the different trajectories that connect hospital admission $A$ to the absorbing state, $D$, $H$ or $L$,  that determine  the patient's condition on discharge from hospital.
 \begin{align}
 \theta_{A \bigcdot D}= & \,\,\theta_{A\, I\,W2\,D}+ \theta_{A\,I\,D}  + \theta_{A\,W1\,I\, W2\, D} +\theta_{A\,W1\,I \,D} +\theta_{A\,W1\,D}
 \nonumber\\
 = &  \,\,\theta_{A\,I}\,  \theta_{I\,W2}\, \theta_{W2\,D} + \theta_{A\,I}\,\theta_{I\,D} + \theta_{A\,W1} \,\theta_{W1\,I} \,\theta_{I\, W2} \,\theta_{W2\, D} \nonumber\\
  &  +\,\, \theta_{A\,W1} \,\theta_{W1\,I}\, \theta_{I \,D} + \theta_{A\,W1}\,\theta_{W1\,D}. \nonumber \\
 \theta_{A\bigcdot H}= & \, \theta_{A\,W1\,H}+ \theta_{A\,W1\,I\,W2\,H} + \theta_{A\,I\,W2\,H} = \theta_{A\,W1}\,\theta_{W1\,H}\nonumber\\
 & + \,\,  \theta_{A\,W1} \, \theta_{W1\,I}  \,\theta_{I\,W2} \, \theta_{W2\,H} + \theta_{A\,I}\,\theta_{I\,W2}\,\theta_{W2\,H}. \nonumber\\
 \theta_{A   \bigcdot   L}= & \,\theta_{A\,W1\,L} + \theta_{A\,W1\,I\,W2\,L} + \theta_{A\,I\,W2\,L} = \theta_{A\,W1}\, \theta_{W1\,L} \nonumber\\  & + \,\,\theta_{A\,W1} \, \theta_{W1\,I}  \,\theta_{I\,W2} \, \theta_{W2\,L} + \theta_{A\,I} \, \theta_{I\,W2} \, \theta_{W2\,L}. \label{eqn:3}
 \end{align}

 The posterior distribution for these probabilities, $\pi(\theta_{A \bigcdot D} \mid \mathcal{D})$, $\pi( \theta_{A \bigcdot H} \mid \mathcal{D})$, and $\pi( \theta_{A \bigcdot L} \mid \mathcal{D})$, is not analytical but, as mentioned above, it can be approximated via simulation by generating approximate samples of the  posterior distribution of each direct transition probability following (\ref{eqn:1}).

\begin{figure}[H]
\begin{center}
\includegraphics[width=12.5cm, height=5cm]{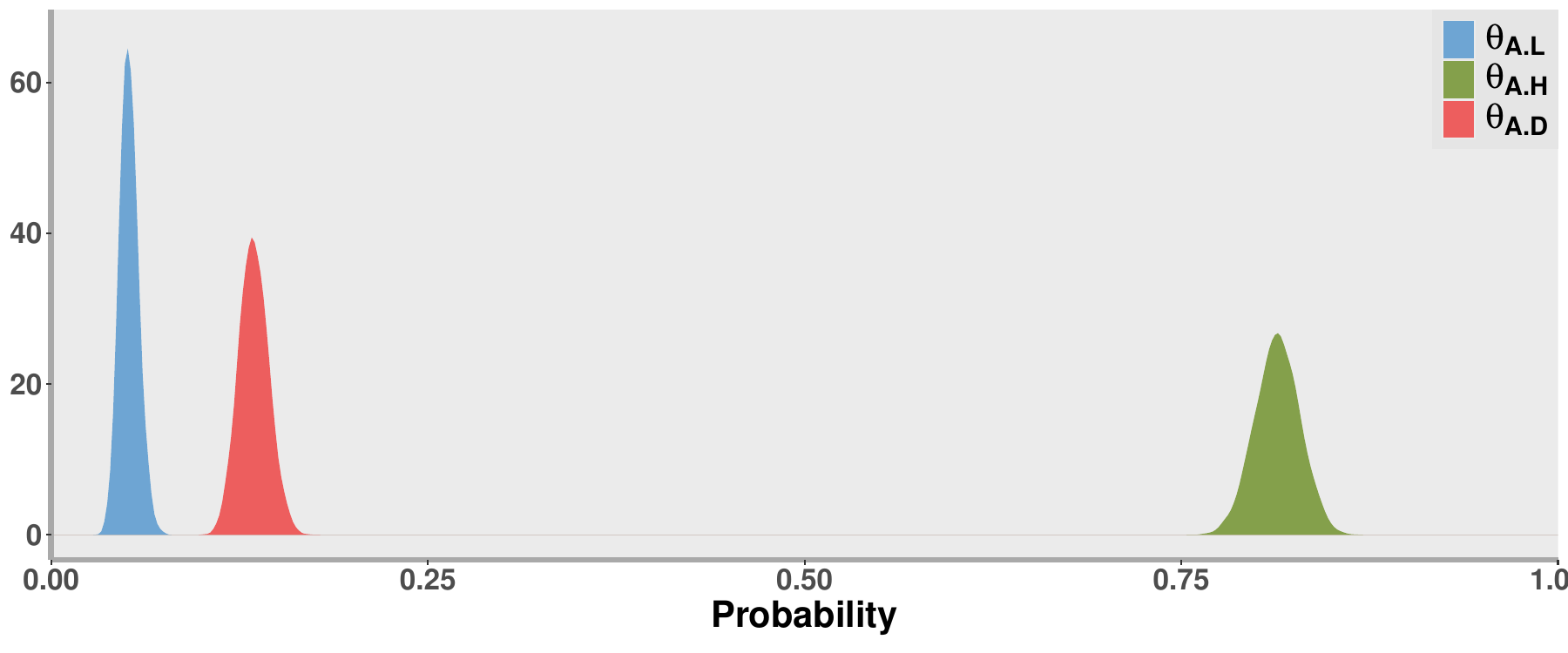}
    \caption{Posterior distribution for the probability of  dying in hospital. being sent to a long-term institution or being discharged cured and sent home.  }\label{fig:6}
\end{center}
\end{figure}

Figure \ref{fig:6} displays the posterior distribution for the probability that a patient hospitalized with severe influenza will die in the hospital, be cured, discharged and sent home, or be transferred to a long-term care unit.

Posterior expectation and  95$\%$ credible intervals for these probabilities are:
    \begin{align*}
    \text{E}(\theta_{A \bigcdot  L} \mid \mathcal D)= 0.052,& \,\,\, 0.95\,\text{CI}(\theta_{A\bigcdot  L} \mid \mathcal D)  = (0.040, 0.065)\nonumber\\
    \text{E}(\theta_{A\bigcdot  H} \mid \mathcal D)= 0.814, &\,\,\,\, 0.95\,\text{CI}(\theta_{A\bigcdot  H} \mid \mathcal D)= (0.784, 0.843),  \nonumber \\
    \text{E}(\theta_{A\bigcdot  D} \mid \mathcal D)= 0.134, &\,\,\,\, 0.95\,\text{CI}(\theta_{A\bigcdot  D} \mid \mathcal D)= (0.116, 0.155),  \nonumber
    \end{align*}

About 81$\%$ of patients who are hospitalised due to severe influenza will be discharged cured and sent home, about 13$\%$ will die and about 5$\%$  will be sent to a long-term care facility.  As the credible intervals above and the posterior distributions  for $\theta_{A \bigcdot  L}$, $\theta_{A \bigcdot  H} $, and $\theta_{A \bigcdot  D} $ shown    in Figure \ref{fig:6} indicate, the uncertainty associated with each of these estimates is quite small, especially that associated with dying in hospital or being sent to a long-term institution.

\subsection{Probability that a patient who has died, or has been discharged and sent home or has been sent to a long-term institution has spent time in the
ICU.}
 In our study, it may be interesting to simulate the posterior distribution of some inverse probabilities, such as the probability that a patient who died in the
 hospital, was sent to a long-term care facility or was discharged cured had
 previously been in the ICU, $\theta_{D \bigcdot I}$, $\theta_{L\bigcdot  I}$,
or $\theta_{H im\bigcdot  I}$. We start with the posterior distribution for  $\theta_{D\bigcdot  I}$. Following (\ref{eqn:2})

$$\theta_{D\bigcdot  I}= \frac{\theta_{I\bigcdot D}\,\theta_{D}}{\theta_{I}},$$

\noindent where $\theta_D$ ($\theta_I$) is the probability that a hospitalized
patient dies in hospital (enters the UCI)   which we have previously
represented as $\theta_{A \bigcdot D}$ ($\theta_{A \bigcdot I}$),  with
\begin{align}
\theta_{I\bigcdot  D}= & \, \theta_{I\,D} + \theta_{I\,W2\,D}= \theta_{I\,D} + \theta_{I\,W2}\, \theta_{W2\,D}, \nonumber \\
\theta_{A\bigcdot  I}= & \,\theta_{A\,I} + \theta_{A\,W1\,I}= \theta_{A\,I} + \theta_{A\,W1}\, \theta_{W1\,I}, \nonumber
\end{align}
and  $\theta_{A \bigcdot D}$  expressed in (\ref{eqn:3}) in terms of transition probabilities between contiguous states. This   procedure can be also followed to simulate the posterior distribution of the rest of inverse probabilities, $\theta_{L\bigcdot  I}$ and
 $\theta_{H \bigcdot  I}$.

\begin{figure}[H]
\begin{center}
\includegraphics[width=12.5cm, height=5cm]{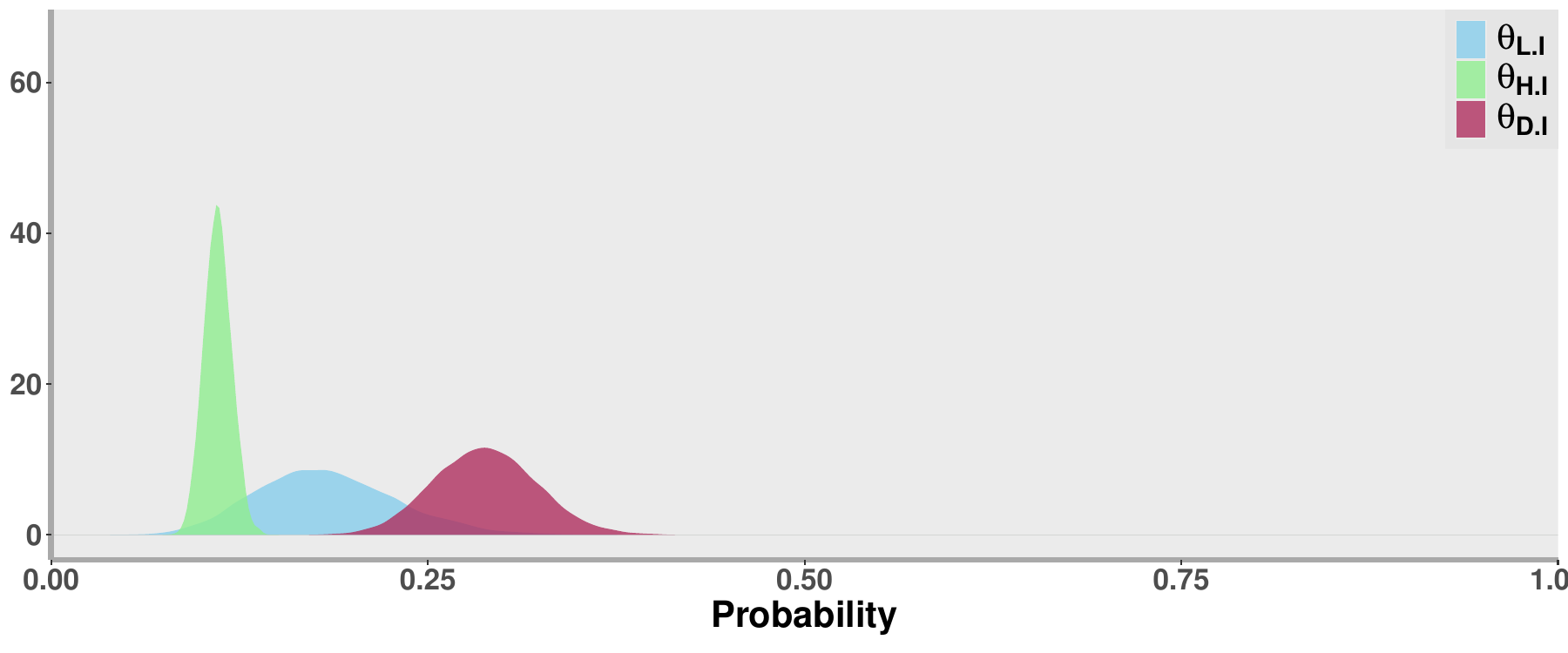}
    \caption{Probability that a patient who has died, or has been
discharged and sent home or has been sent to a long-
term institution has spent time in the ICU.  }\label{fig:7}
\end{center}
\end{figure}

Figure \ref{fig:7} shows the posterior distribution of the three previous probabilities. We can see that people who were discharged cured from the hospital are the least likely to have spent time in the ICU ($\text{E}(\theta_{H\bigcdot  I} \mid \mathcal D)= 0.110$), followed by patients who were sent to a long-stay ward ($\text{E}(\theta_{L \bigcdot  I} \mid \mathcal D)= 0.183$), and finally people who died in hospital ($\text{E}(\theta_{D\bigcdot  I} \mid \mathcal D)= 0.288$). We can also observe that the distribution associated with leaving the hospital cured is the one with the least uncertainty, probably because the corresponding sample size is larger. In fact, the posterior 95$\%$ credible intervals are $\text{CI}(\theta_{H\bigcdot  I} \mid \mathcal D)= (0.093, 0.128)$, $\text{CI}(\theta_{L\bigcdot  I} \mid \mathcal D)  = (0.100, 0.280)$, and $\text{CI}(\theta_{D\bigcdot  I} \mid \mathcal D)= (0.224, 0.357)$.

\section*{Conclusions}
Probabilistic DAGs are very helpful representations of complex environments with stochastic dependencies. Bayesian inference in DAGS defined by random events is a powerful framework for understanding and assessing the prevalence and uncertainty associated with the different  trajectories in the system.

The representation of the evolution of patients admitted to hospital as a consequence of severe influenza through a DAG and the subsequent statistical analysis provides valuable clinical information on the severity of the disease as well as on the utilisation of healthcare resources.  This information is key to hospital resource planning because it helps identify the human, material and financial resources needed to ensure quality and efficient hospital care.

Our work provides a flexible framework that allows  the inclusion of potentially relevant additional information in terms of demographic (sex, age) or clinical (comorbidities) covariates, as well as other types of information such as length of stay in the different services, or even considering the potential variability between the different hospitals participating in the study.\vspace*{0.2cm}\\

\noindent {\Large \textbf{Acknowledgements} } \vspace*{0.2cm}\\
This paper is partially supported by  the project PID2023-148158OB-I00, funded by Ministerio de Ciencia, Innovación y Universidades (Spain) and  the project PID2022-136455NB-I00, funded by Ministerio de Ciencia, Innovación y Universidades of Spain (MCIN/AEI/10.13039/501100011033/ FEDER, UE) and the European Regional Development Fund.

\bibliographystyle{chicago}

\end{document}